\documentclass[twocolumn,showpacs,preprintnumbers,amsmath,amssymb]{revtex4}
\usepackage{lipsum}
\usepackage{graphicx}
\usepackage{dcolumn}
\usepackage{bm}
\usepackage[final]{pdfpages}
\begin{document}
\title{Fermi surface of the most dilute superconductor}
\author{Xiao Lin$^{1,2}$,  Zengwei Zhu$^{1}$, Beno\^{\i}t Fauqu\'e$^{1}$,  and Kamran Behnia$^{1}$\email{kamran.behnia@espci.fr}}
\affiliation{$^{1}$LPEM (UPMC-CNRS), Ecole Sup\'erieure de Physique et de Chimie Industrielles, 75005 Paris, France \\
$^{2}$ Department of Physics, Zhejiang University, Hangzhou, 310027, China}

\date{January 10, 2013}

\begin{abstract}
\textbf{ The origin of superconductivity in bulk SrTiO$_{3}$ is a mystery, since the non-monotonous variation of the critical transition with carrier concentration defies the expectations of the crudest version of the BCS theory. Here, employing the Nernst effect, an extremely sensitive probe of tiny bulk Fermi surfaces, we show that down to concentrations as low as $5.5 \times 10^{17}cm^{-3}$, the system has both a sharp Fermi surface and a superconducting ground state. The most dilute superconductor currently known has therefore a metallic normal state with a Fermi energy  as little as 1.1 meV on top of a band gap as large as 3 eV. Occurrence of a superconducting instability in an extremely small, single-component and barely anisotropic Fermi surface implies strong constraints for the identification of the pairing mechanism.}
\end{abstract}
\maketitle

\section{Introduction}
SrTiO$_{3}$ is a large-gap transparent insulator, which upon the introduction of  n-type carriers, undergoes a superconducting transition below 1 K. Discovered in 1964\cite{schooley}, it has been the first member of a loose family of ``semiconducting superconductors''\cite{hulm}, which now includes column-IV elements \cite{blase}. During the last decade, attention has focused on the interface between SrTiO$_{3}$ and other insulators\cite{ohtomo} or vacuum\cite{santander}, which is a two-dimensional metal with a superconducting ground state\cite{reyren,biscaras}.

\begin{figure}\resizebox{!}{0.95\textwidth}
{\includegraphics{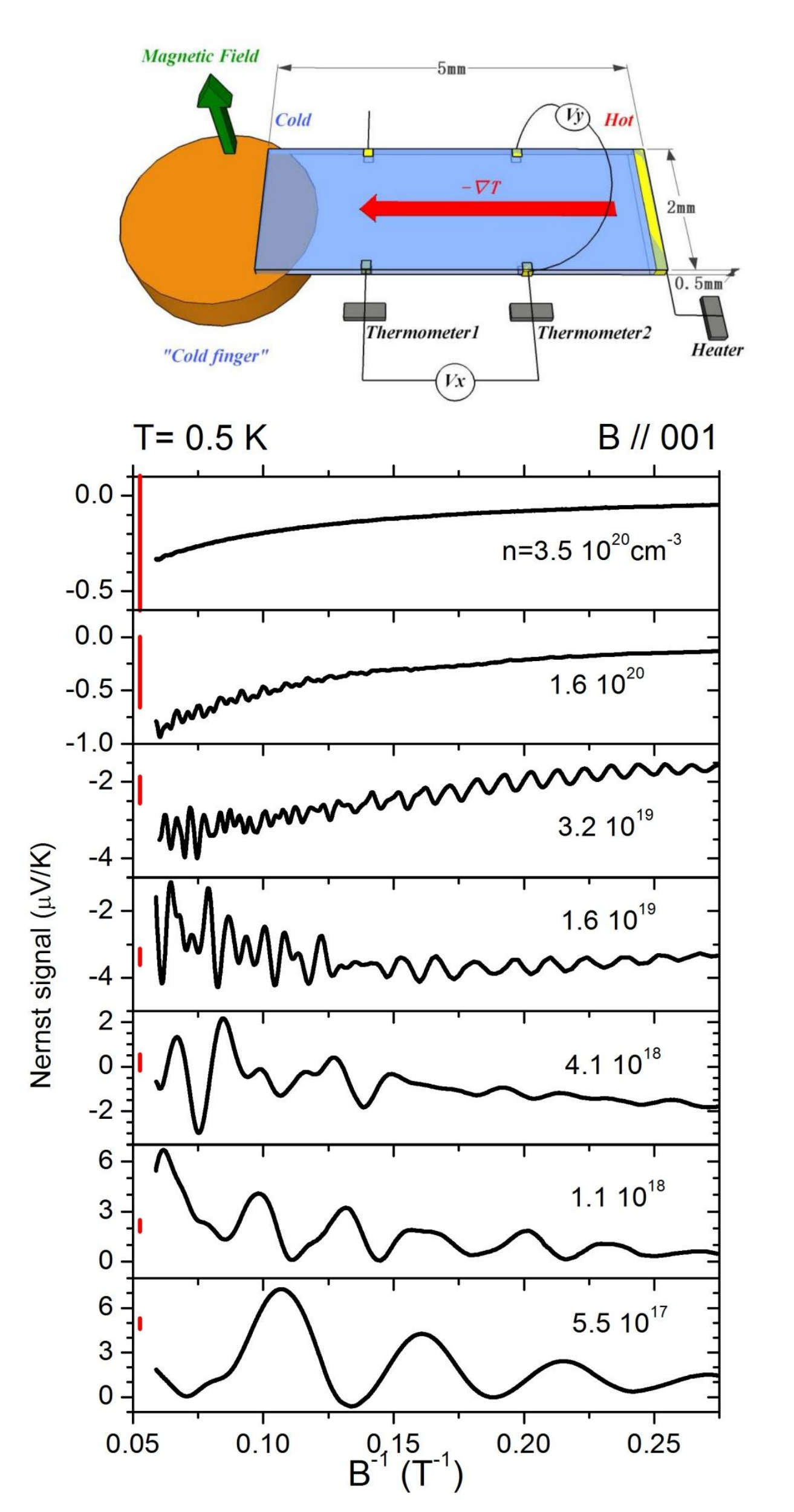}}\caption{ \textbf{Top:} The expeimental set-up for measuring Nernst and Seebeck coefficients of bulk chemically-doped SrTiO$_{3}$. \textbf{Bottom:} As the carrier density is reduced in SrTiO$_{3}$, the Nernst signal shows oscillations with  larger amplitude and longer periodicity. In each panel, the vertical red bar represents a constant scale of 0.7 $\mu V/K$.}
\end{figure}

According to Mott, in a doped semiconductor, when the average distance between the dopants ($d= n^{-1/3}$) becomes a sizeable fraction of the effective Bohr radius, $a_{B}^{*}$, metal-insulator transition occurs. Quantitatively, this Mott criterion for the critical concentration is expressed as $n_{c}^{1/3}a_{B}^{*}=0.26$ and has been observed to hold in a wide range of semiconductors\cite{edwards}.

\begin{figure}
\resizebox{!}{0.7\textwidth}{\includegraphics[width=10cm]{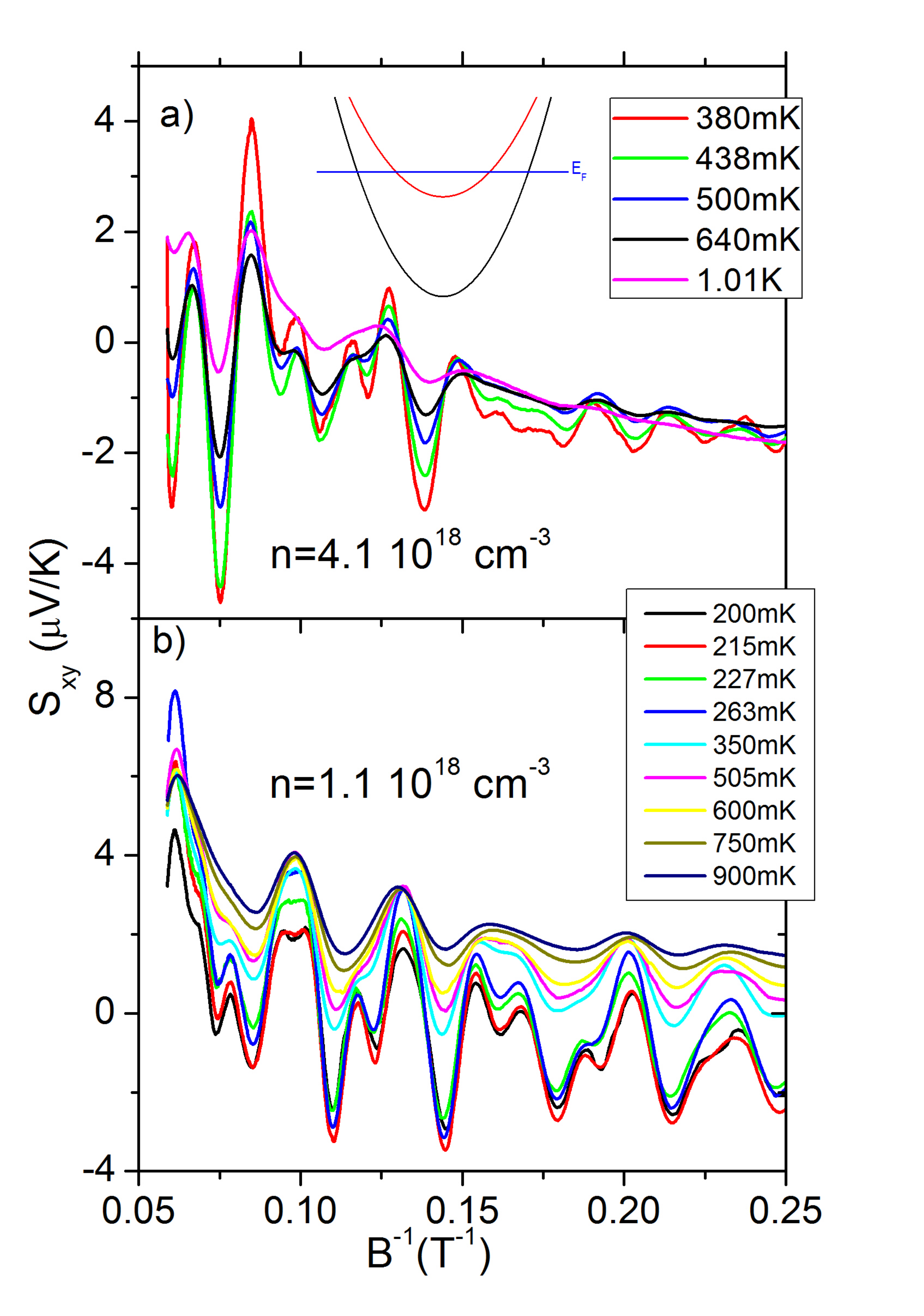}}
\caption{Nernst signal as a function of the inverse of magnetic field for two carrier concentrations. Quantum oscillations display a complex structure with multiple periodicities. The inset shows a sketch of the position of Fermi energy relative to the two lowest bands at $\Gamma$-point.}
\end{figure}

Insulating SrTiO$_{3}$ is dubbed a quantum paraelectric. At low temperature, its static dielectric constant becomes 4 orders of magnitude larger than vacuum\cite{muller}. Since the effective Bohr radius is  proportional to the dielectric constant, this implies a very long Bohr radius. Therefore, following the Mott criterion, the  critical density for metal-insulator transition is expected to be much lower than in ordinary semiconductors and indeed, doped SrTiO$_{3}$ displays a finite zero-temperature conductivity down to carrier concentrations as low as $n=8\times 10^{15}cm^{-3}$ \cite{spinelli}. This is orders of magnitude lower than the the threshold of metallicity in silicon ($3.5\times 10^{18}cm^{-3}$) or in germanium ($3.5\times 10^{17}cm^{-3}$)\cite{edwards}.

\begin{figure*}\centering
\resizebox{!}{0.7\textwidth}{\includegraphics{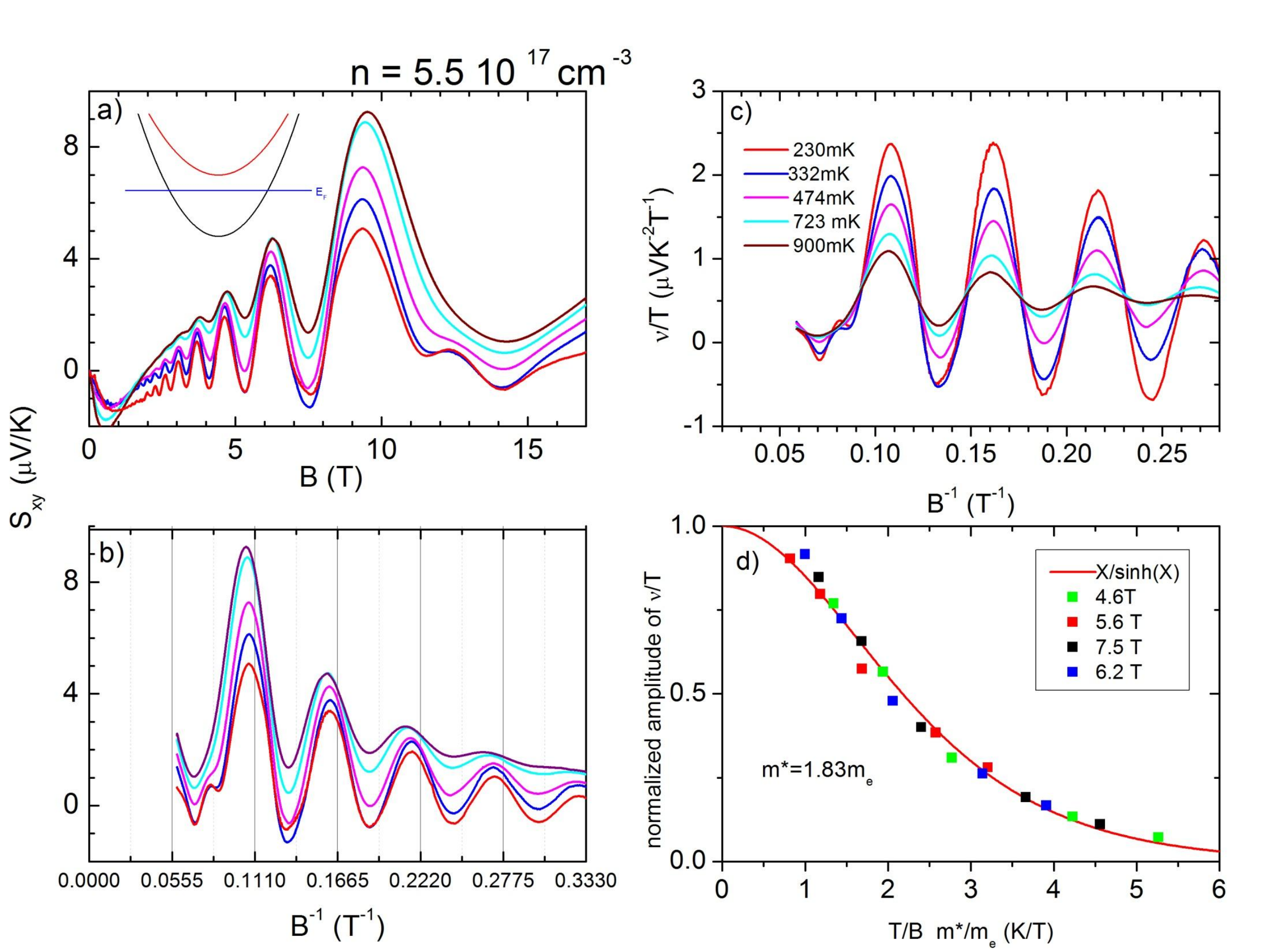}} \caption{Nernst quantum oscillations as a function of magnetic field \textbf{(a)} and the inverse of magnetic field \textbf{(b)} at different temperatures for the sample with a carrier density of $5.5 \times 10^{17} cm^{-3}$. There is a single periodicity with the lowest peak displaying Zeeman splitting. \textbf{c)}The variation of Nernst coefficient ($\nu=S_{xy}/B$ divided by temperature with the inverse of magnetic field at different temperatures. \textbf{d)}The amplitude of oscillations in $\nu/T$, for different magnetic fields and temperatures. Solid line represents the damping expected in the Lifshitz-Kosevitch theory for an effective mass of $1.83m_{e}$.}
\end{figure*}

At still higher doping levels, the system becomes a superconductor. This is achieved by $n-$doping the system by either of the three possible routes: substituting titanium with niobium\cite{schooley2,koonce,binnig}, or strontium with lanthanum\cite{suzuki} or removing oxygen\cite{schooley2}. Intriguingly, the superconducting ground state is restricted to a limited doping window\cite{schooley2,koonce,binnig,suzuki}. Thanks to the remarkably high mobility of electrons, another consequence of the large dielectric constant\cite{spinelli}, quantum oscillations are observable and have been reported both in the bulk\cite{frederikse,gregory} and in the two-dimensional\cite{kozuka,benshalom,caviglia,kim} samples.

The main subject of the present work is to yield a quantitative description of the emerging Fermi surface and to clarify its relationship with the superconducting ground state. Our work addresses two hitherto unanswered questions: i) Is there a threshold in carrier concentration for the emergence of superconductivity? ii) Has the normal state of such a low-density superconductor a well-defined Fermi surface or is an impurity-band metal? We provide answers to these questions by a study of the low-temperature Nernst effect (a very sensitive probe of tiny Fermi surfaces\cite{behnia1,zhu,fauque}) in both oxygen-reduced and Nb-doped SrTiO$_{3}$ across a wide (i.e. three-orders-of-magnitude) window  of carrier density.

We find that superconductivity persists down to a carrier concentration significantly lower than what was previously believed. This firmly establishes $n-$doped SrTiO$_{3}$ as the most dilute known superconductor with a carrier density as low as $5.5\times10^{17}cm^{-3}$, which corresponds to the removal of one oxygen atom out of $10^{5}$. At this carrier concentration, giant Nernst quantum oscillations with a single frequency are observed. The superconductivity persists even in presence of a single, barely filled and almost isotropic band.  We will argue that this context is radically different from what one finds in conventional phonon-mediated BCS superconductors. Thus, more than four decades after its discovery, SrTiO$_{3}$ emerges out of this study as a candidate for unconventional superconductivity.

\section{Experimental}

The study was carried out on bulk commercial STO single crystals.  Oxygen-deficient samples were obtained by annealing nominally stochimoteric samples . Hall and longitudinal resistivity of the samples were measured and the carrier dependence of mobility and room-temperature resistivity were found to be compatible with early reports by Fredrikse and co-workers\cite{Frederikse1,Frederikse2} and recent work by Spinelli and co-workers\cite{spinelli}. In order to obtain ohmic contacts, gold was evaporated on the samples and heated up to 550 degrees. Typical contact resistance was a few Ohms at room temperature and well below 1 Ohm at low temperature. A one-heater-two-thermometers set-up was used to measure all transverse and longitudinal electric and thermoelectric coefficients (Resistivity, Hall, Seebeck and Nernst). A sketch of the set-up is shown in the upper panel of Fig. 1. At low temperatures, a noise level of 1nV was achieved. The accuracy of thermal gradient across the sample was checked by retrieving the Wiedemann-Franz relation between thermal and electric conductance in each sample. Details on sample preparation and characterization are given in the supplement.

\section{Results and discussion}
Fig. 1 presents the variation of the Nernst signal (S$_{xy}$=E$_{y}$/-$\nabla_{x}$T) as a function of the inverse of the magnetic field for seven samples with different carrier concentrations, kept at the same temperature (0.5 K). As seen in the figure, quantum oscillations are barely detectable in the sample with the highest carrier density. As $n$ decreases, the amplitude of oscillations grows and their frequency shrinks. Giant oscillations of the Nernst effect with the approach of the quantum limit were previously observed in semimetallic bismuth\cite{behnia1} and graphite\cite{zhu}, as well as doped Bi$_{2}$Se$_{3}$\cite{fauque}. A property these three systems share with lightly-doped SrTiO$_{3}$ is that their Fermi surface is an extremely small portion of the Brillouin zone. A 10 T magnetic field truncates such a tiny Fermi surface in to a few Landau tubes. In such a context, each time a squeezed Landau tube leaves the Fermi surface, the Nernst signal peaks. The Nernst quantum oscillations are concomitant with Shubnikov-de Haas osillations.  For the lowest Landau indexes, however, while the oscillating component of resistivity is a small fraction of the overall signal,  the oscillating part of the Nernst coefficient dominates the monotonous background\cite{zhu}. This makes the analysis of the Nernst data straightforward.

At low magnetic fields, the monotonous Nernst signal is expected to be affected by fluctuating superconductivity. In a superconductor, fluctuations of the superconducting order parameter can generate a Nernst signal above T$_{c}$. Theory has quantified the magnitude of the signal due to the Gaussian fluctuations of the superconducting order parameter in both two and three dimensions\cite{ussishkin}. On the other hand, the contribution of normal quasi-particles to the Nernst signal is proportional to the ratio of their mobility to their Fermi energy\cite{behnia3}. In a dirty high-density two-dimensional superconductor with low electron mobility and large Fermi energy, the quasi-particle contribution is small and the superconducting Nernst signal is detectable in a large temperature window above T$_{c}$\cite{pourret}. In slightly-doped bulk SrTiO$_{3}$, in contrast, the electron mobility is large, the Fermi energy is small and the Nernst signal is therefore dominated by the quasi-particle contribution [See the supplement for a comparison of the order of magnitude of quasi-particle and superconducting contributions to the Nernst signal].

As seen in Fig. 1, in the intermediate doping range, the oscillations display a complex structure and several frequencies are detectable.
For samples with larger carrier densities ($ n\geq 1.05 \times 10^{18}cm^{-3}$), the spectrum of oscillations is indicative of the presence of more than one component of the Fermi surface. For the lowest doping level, the structure becomes remarkably simpler.

Fig.2 displays detailed data at different temperatures for two low-density samples. Quantum oscillations show a complex structure. At $n= 4.1\times10^{18} cm^{-3}$ more than one frequency are clearly detectable, indicating that the Fermi level is above the bottom of the second band and there are two occupied bands. At $n= 1.05\times10^{18} cm^{-3}$, we detect two sets of peaks with slightly different frequencies (26.8 T and 28.3 T). This may correspond to two different bands or a single  spin-split band.

One major result reported here is the unambiguous detection of single-frequency quantum oscillations at a carrier density of $n= 5.5\times10^{17} cm^{-3}$. The data for this sample are shown in Fig. 3. As seen in the figure, there is only a single periodicity for all temperatures. Moreover, the Nernst peak for the lowest detected Landau level is split at low enough temperatures. The data immediately settles the periodicity ($0.0555 T^{-1}$) corresponding to a frequency of 18.2 T. The distance between the two spin-split-peaks quantifies the relative magnitudes of the Zeeman and cyclotron energies. This is the first time that the Zeeman splitting has been clearly detected in the system.
\begin{figure}
\resizebox{!}{0.7\textwidth}{\includegraphics{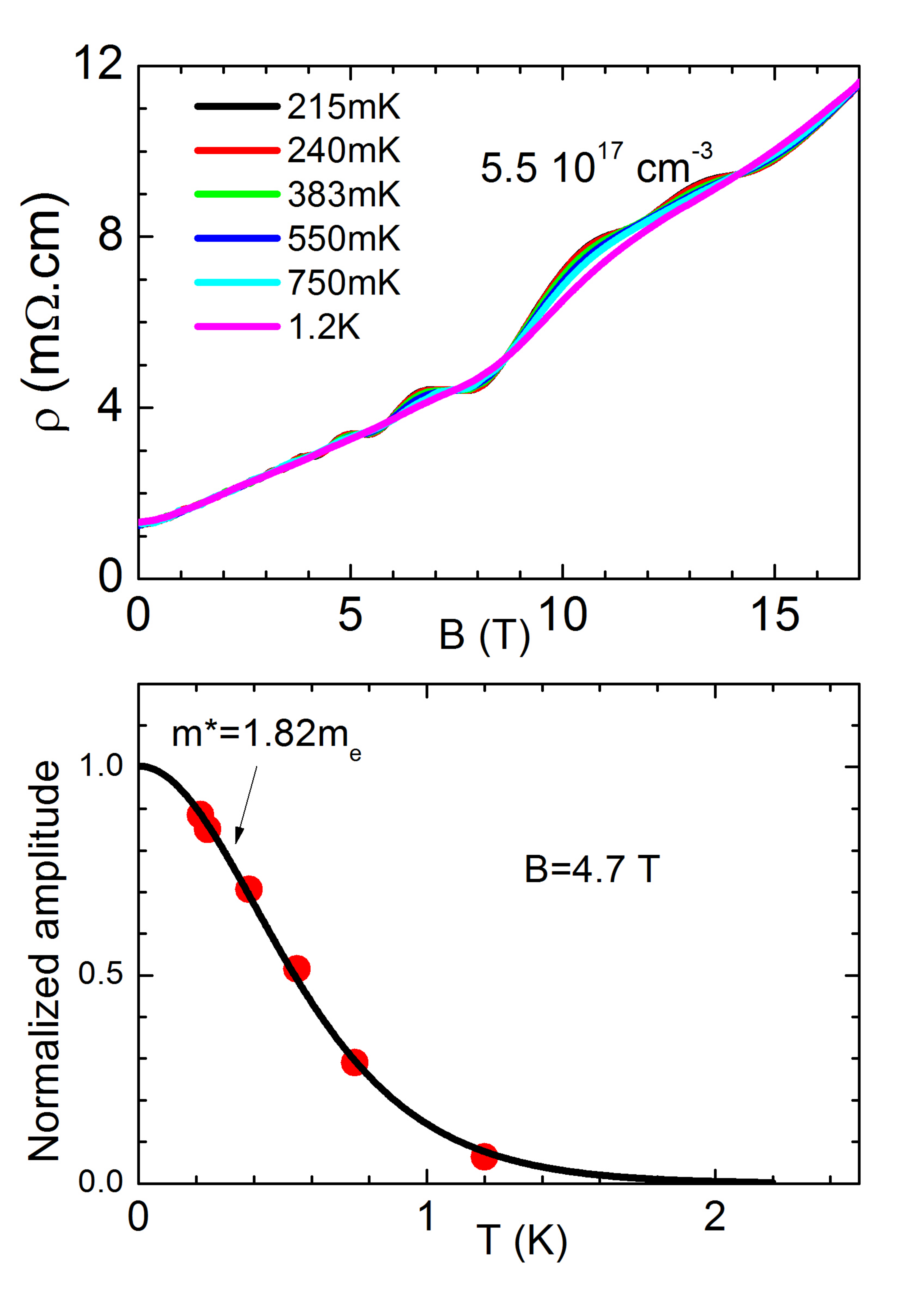}}
\caption{\textbf{Top}: Shubnikov-de Haas effect is visible in the magnetoresistance  of the lowest doped SrTiO$_{3}$ sample. \textbf{Bottom:} The effective mass obtained from resistivity data is close to the one obtained by the analysis of the Nernst data.}
\end{figure}

Let us compare these findings with the conclusions  of recent \emph{ab initio} band calculations\cite{vandermarel}. As the insulator is doped by n-type carriers, the first available band to be filled is a threefold degenerate one associated with Ti \emph{3d} orbital. The threefold degeneracy is lifted first by spin-orbit coupling and then by the crystal electric field. Filling these bands successively generates three interpenetrating ellipsoids at the center of the Brillouin zone. According to these calculations, at a critical doping of $x_{c1}= 4\times 10^{-5}$ carrier per formula unit\cite{vandermarel} the second band starts to be filled. This corresponds to a carrier density of $n_{c1}=6.8\times 10^{17} cm^{-3}$, slightly larger than our most dilute sample. Therefore, there is a fair agreement between experiment and theory.

\begin{figure*}\centering\resizebox{!}{0.7\textwidth}
{\includegraphics{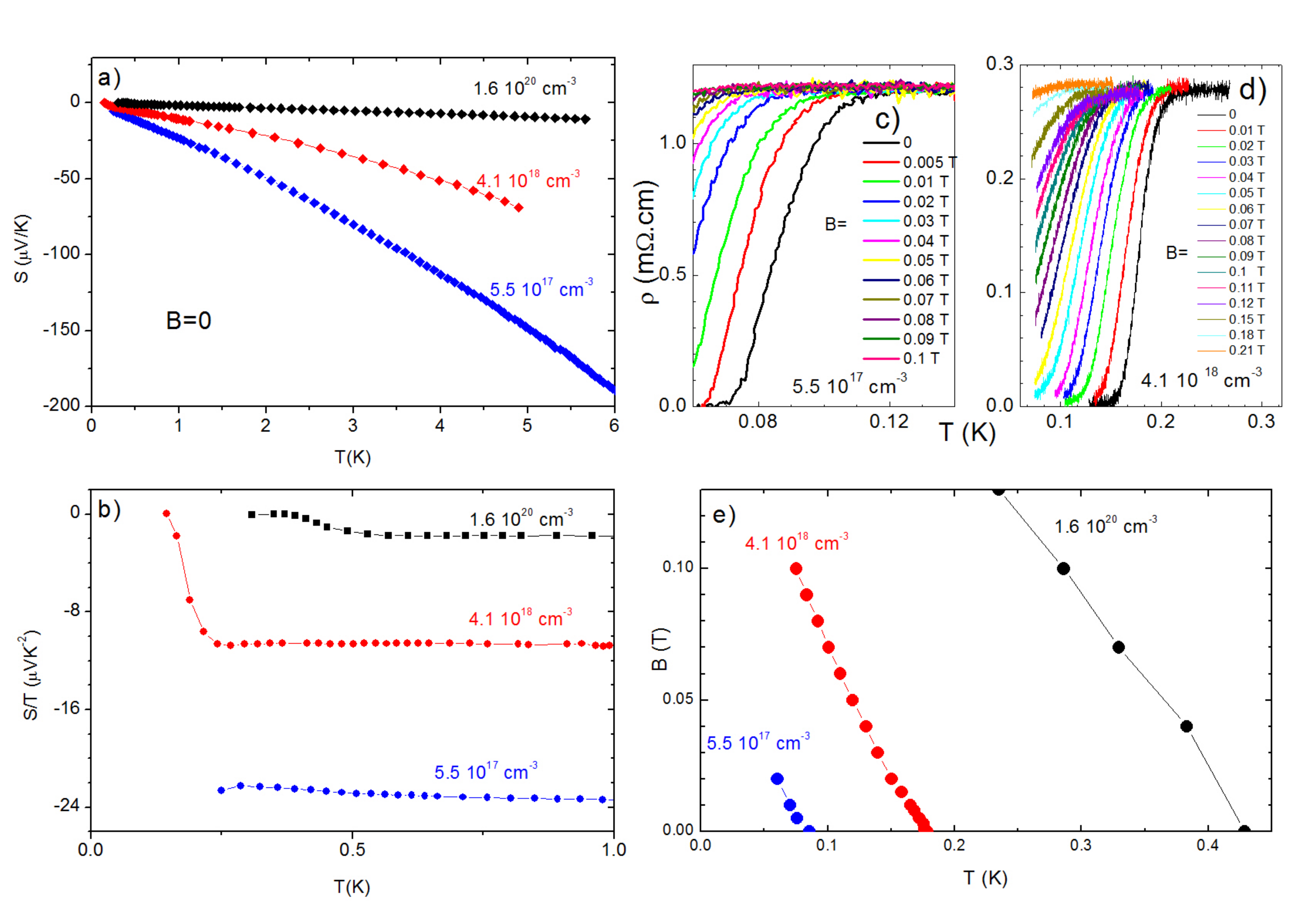}}
\caption{\textbf{a)} The temperature dependence of the zero-field Seebeck coefficient for three different carrier concentrations.  \textbf{b)}Temperature dependence of $S/T$. As seen in the figure,  the diffusive component of thermopower becomes larger with underdoping, indicative of a decrease in the Fermi energy. \textbf{c and d)} Superconducting resistive transition for two different carrier concentrations in presence of magnetic field. \textbf{e)} Temperature-dependence of the upper critical field  near  T$_{c}$ (defined as the temperature at which resistivity drops by half) for three carrier concentrations. }
\end{figure*}

According to the Onsager relation, the frequency of quantum oscillations, F, is set by the extremal cross section of the Fermi surface,$A_{k}$: $F=(\frac{\hbar}{2\pi e})A_{k}$. A frequency of 18.2 T implies a cross section of $A=0.18 nm^{-2}$. Now, assuming a spherical Fermi surface, this gives a carrier density of $4.6\times10^{17}cm^{-3}$, very close to what is given by measuring the Hall resistivity ($n=\frac{1}{eR_{H}}=5.5\times10^{17}cm^{-3}$). We conclude that at this concentration, the Fermi surface is a slightly anisotropic sphere. This is again in agreement with the expectations of the band calculations, which do not find sizeable anisotropy in dispersion along [100], [101] and [110] for $n < n_{c1}$.

In the Lifshitz-Kosevitch (LK) theory, the thermal smearing of quantum oscillations is set by the magnitude of cyclotron mass, $m^{*}$. As the temperature increases, the ratio of thermal energy to the cyclotron energy increases.Therefore the amplitude of  oscillations decreases following $\frac{X}{sinh X }$ dependence. Here, $X$ is a dimensionless temperature normalized by magnetic field and effective mass: $X=(\frac{2\pi^{2}k_{B}}{\hbar e})(\frac{m^{*}T}{B})$. As in the case of Bi$_{2}$Se$_{3}$\cite{fauque}, we found that employing this formalism to oscillations of the Nernst coefficient divided by temperature ($\nu/T$) (See Fig. 3c) leads to an accurate determination of the cyclotron mass. Fig. 3d, compares the experimentally-resolved amplitudes at different field and temperatures with the expectations of LK equation assuming $m^{*}=1.83m_{e}$. Fig.4 displays the resistivity and its quantum oscillations in the same sample. As seen in the bottom panel of the figure, a fit to our Shubnikov- de Haas data yields a quasi-identical value for the effective mass of $m^{*}=(1.82\pm 0.05)m_{e}$ .

The effective mass obtained here is in fair agreement with what was estimated by measuring the plasma frequency ($m^{*}=(2\pm 0.3) m_{e})$\cite{vanmechelen}. Note, however, that the latter measurements are performed at higher concentrations where several bands with possibly different masses are present. The cyclotron mass obtained here is 2.5 times larger than the the band mass(0.7 $m_{e}$\cite{vanmechelen,vandermarel}). This is presumably due to a combination of electron-phonon and electron-electron interactions.

The Fermi temperature of this dilute liquid of electrons can be estimated in two distinct ways. First, the  magnitude of cyclotron mass together with the size of the Fermi wave-vector $k_{F}$ obtained from the frequency of the quantum oscillations can be plugged to:
\begin{equation}\label{1}
k_{B}T_{F}=\frac{\hbar^{2}k_{F}^{2}}{2m^{*}}
\end{equation}
to obtain the Fermi temperature. This yields  a Fermi temperature of 13.5 K.

Another route to estimate the Fermi temperature is to put under scrutiny the magnitude of the Seebeck coefficient. In a wide variety of correlated metals, the slope of the Seebeck coefficient in the zero-temperature limit is inversely proportional to the Fermi temperature as expected in the semi-classical transport theory for a Fermi liquid\cite{behnia2}. In doped SrTiO$_{3}$, the magnitude of the Seebeck coefficient in the intermediate temperature range is known to be remarkably large\cite{okuda}. Panels a and b in Fig. 5 display  the evolution of low-temperature Seebeck coefficient in SrTiO$_{3}$ with doping. As seen in the figure, as the carrier density lowers,  the diffusive component of the Seebeck coefficient increases. In the sample with the lowest carrier concentration ($n=5.5\times10^{17}cm^{-3}$) , the magnitude of $S/T$ becomes as large as $-22\mu VK^{-2}$. Assuming an energy-independent mean-free-path and a spherical Fermi surface, a Fermi temperature of T$_{F}= 12.9 K$ is obtained using the equation:
\begin{equation}\label{1}
|\frac{S}{T}|=\frac{\pi^{2}}{3}\frac{k_{B}}{e}\frac{1}{T_{F}}
\end{equation}

Thus, data obtained from two independent probes converge to a Fermi temperature as low as 13 K. The energy gap between valence and conduction bands in SrTiO$_{3}$ is as large as 3eV. By removing one oxygen out of $10^5$ (assuming that each oxygen vacancy liberates two potentially mobile electrons), one can create a metal with a chemical potential as small as 1.1 meV on the top of this gap. It is remarkable that this can be achieved in spite of unavoidable inhomogeneities in dopant distribution. Long-range screening, which damps local band-bending effects appears to be the key factor here. This is the first major result coming out of this investigation.

This dilute liquid of electrons becomes a superconductor below a critical temperature of T$_{c}$=86 mK. This second unexpected result is illustrated in Fig. 5c, which shows the resistive transition. The width of transition (defined as the difference between the temperatures of 10 percent and 90 percent drops in resistivity) is 25 mK. This is to be compared with a width of 30 mK in a sample with a T$_{c}$ of 180 mK and a carrier density of $4.1 \times 10^{18} cm^{-3}$  (Fig. 5d). As seen in the figure, the superconducting transition is easily suppressed by the application of a small magnetic field.

\begin{figure}
\resizebox{!}{0.7\textwidth}{\includegraphics{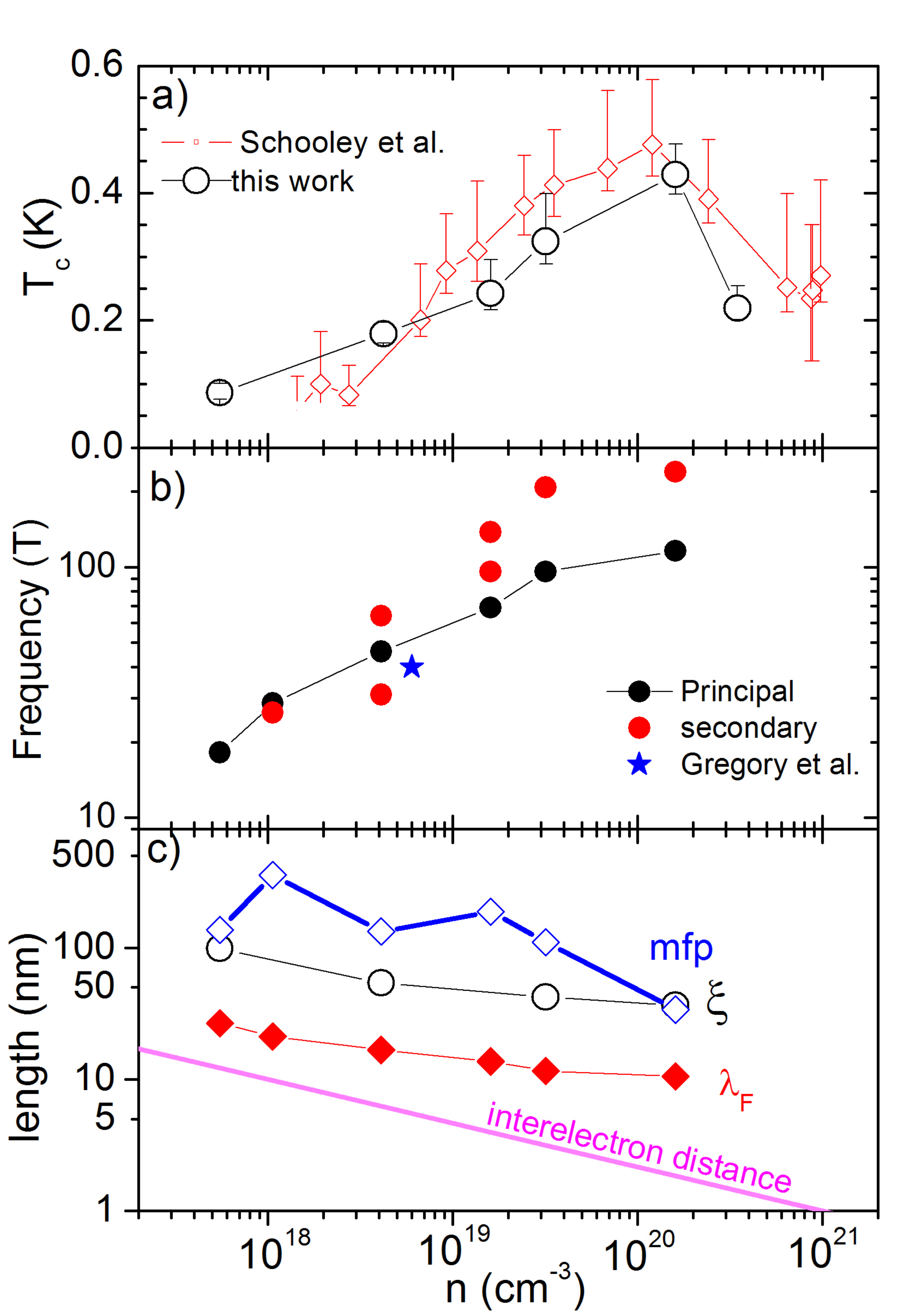}}
\caption{\textbf{a)} Variation of the critical temperature with carrier concentration according to this work and ref.\cite{schooley2}. The width of superconducting transition in each case is represented by error bars. \textbf{ b)} Evolution of detected frequencies as a function of doping. For carrier concentrations larger than $10^{18} cm^{-3}$, oscillations display more than one frequency. The frequency and carrier density reported in ref.\cite{gregory} is also shown. \textbf{c)} Variation of four relevant length scales, namely the electronic mean-free-path, the average Fermi wavelength, $\lambda_{F}$ of electrons of the main band, the superconducting coherence length, $\xi$ and the interelectron distance.}
\end{figure}

Fig. 6 details the evolution of superconducting and normal state properties with doping. As seen in the upper panel, the evolution of T$_{c}$ with carrier concentration in our samples is comparable with those reported in ref.\cite{schooley2}, save for the persistence of a long tail on the underdoped side of the phase diagram. The middle panel presents the evolution of frequencies resolved by quantum oscillations with carrier doping. In most samples, in addition to the main frequency, other ones were resolved pointing to the existence of several quantized orbits of electrons as successive bands are filled. For comparison, the panel shows the unique frequency previously reported in the bulk\cite{gregory}. Finally, the lower panel compares the evolution of four different length scales of the system. The average interelectron distance (estimated from carrier density: $d_{ee}=n^{-1/3}$), the average Fermi wavelength (calculated from the Fermi wave-vector, $k_F$ of the main band obtained from the principal frequency:$\lambda_{F}=\frac{2\pi}{k_{F}}$); the mean-free-path (estimated from the measured Hall mobility, $\mu_{H}$ and $k_F$: $\ell_{e}=\frac{\hbar k_{F}}{e} \mu_{H}$) and finally the superconducting coherence length (derived from the slope of the upper critical field: $\xi^{-2}=\frac{2\pi}{\Phi_{0}}0.69T_{c}\frac{dH_{c2}}{dT}|_{T_{c}}$).

Comparing these length scales, we can make several observations. With a \emph{static} dielectric constant as large as $2.5\times10^{4}$\cite{muller}, and an effective mass of $1.8 m_e$, the Bohr radius is of the order of 0.7 $\mu m$. Therefore, the Mott criterion for metallicity ($a_{B}^{*}n^{1/3} >> 0.26)$\cite{edwards} is easily satisfied. Since electrons can travel much further than the interdopant distance, the superconductor remains in the clean limit ($\ell_{e} > \xi$) in our range of study. Finally, the system avoids localisation, because $\ell_{e} >>\lambda_{F}$. As noted previously\cite{vandermarel}, this superconductor stays in the BCS side of the BEC-BCS cross-over. Even for the lowest concentration, the average size of Cooper pairs ($\xi\simeq 100 nm$) is much longer than the interelectron distance ($d_{ee}= 12 nm$).

Superconductivity at such a low carrier density has never been explored in two-dimensional SrTiO$_{3}$, because the electron mean-free-path is much shorter in the latter system. A three-dimensional carrier density of $5.5\times10^{17}cm^{-3}$ corresponds to a 2D carrier density as low as $n_{2D}=6.7\times10^{11}cm^{-2}$. Before attaining such a low carrier density, the superconducting ground state is destroyed by a superconductor-to-insulator transition\cite{caviglia2}, which occurs when $k_{F}\ell \sim 1$.

\section{Conclusion}
Bulk SrTiO$_{3}$ with an oxygen deficiency of 10$^{-5}$ per formula unit is a superconductor. This  carrier density is three to four orders of magnitude lower than what has been reported in conventional ``semiconducting superconductors''. Indeed, boron-doped silicon, diamond or silicon carbide all require a few percent doping to display superconductivity\cite{kriener}. In contrast to SrTiO$_{3}$, the mobility in the latter systems is too low for quantum oscillations and a well-defined Fermi surface has not been observed in the normal state.

SrTiO$_{3}$ is distinct by its single-component, extremely small and barely anisotropic Fermi surface. Such a simple Fermi surface topology is rare among superconductors. It is a well-known fact that noble and alkali metals stand out in the periodical table by their refusal to superconduct. More importantly, the extreme low density generates several peculiar features. The Fermi velocity ($v_{F}=\frac{\hbar k_{F}}{m^{*}}$) becomes as low as 15 km $s^{-1}$, almost as slow as sound velocity. The Fermi temperature is an order of magnitude \emph{lower} than the Debye temperature, a situation never met in a conventional superconductor, but common in a heavy-fermion superconductor\cite{fisk}. The inversion of Debye and Fermi energy scales has profound consequences for the relative weight of Coulomb repulsion and phonon-mediated attraction between electrons\cite{degennes} and is a serious and perhap insurmountable challenge for any theory of superconductivity based on phonon-electron interaction.

The T$_{c}$/T$_{F}$ ratio in dilute SrTiO$_{3}$ is somewhat lower than what is reported than in optimally doped unconventional (cuprate, heavy-fermion and iron-based) superconductors. But, it is an order of magnitude larger than in elemental conventional superconductors and even somewhat larger than in Sr$_{2}$RuO$_4$, a correlated metal believed to be a spin-triplet superconductor\cite{mackenzie}. In the latter case, $T_c$ =1.5 K and each of the three bands  have a Fermi temperature of the order of $T_F \sim1000 K$.

The microscopic mechanism of pair formation in bulk SrTiO$_{3}$ remains a wide open question. Our current knowledge of Fermi surface, single-valley  at the center of the Brillouin zone, leaves no room for intervalley phonons invoked by the oldest theoretical attempt to explain the existence of a superconducting dome in this system\cite{koonce}. On the other hand, the scenario invoking the exchange of plasmons between electrons\cite{takada} appears to be comfortable with the persistence of superconductivity in a dilute one-band context. Little is known about the superconducting order parameter other than the reported observation of a multi-gap superconductivity in the doping range above $10^{19 }cm^{-3}$ \cite{binnig}. Future experiments employing bulk probes such as thermal conductivity or penetration depth are planned to probe the superconducting gap  and find nodes if they exist.

A detailed analysis of quantum oscillations would open a new window to the problem. Since an accurate determination of the Fermi surface topology and effective mass, would lead to the density of states at arbitrary doping. Since the BCS formula ($T_{c}=\Theta exp [\frac{-1}{N(0)V}])$ links T$_{c}$ with an energy scale $\Theta$, an interaction parameter $V$ and the density of states, $N(0)$, competing scenarios can be tested by comparing their expectation for the variation of $V$  and $\Theta$.

\textbf{Acknowledgements-} We acknowledge stimulating discussions with Harold Hwang, Dirk van der Marel, Kazuamasa Miyake and Yasutami Tanaka and are grateful to Young Jun Chang for providing us a number of Nb-doped STO samples. This work is supported by Agence Nationale de la Recherche as part of QUANTHERM and SUPERFIELD projects. X.L. acknowledges a scholarship granted by China Scholarship Council.

\end{document}